\documentclass[prl,superscriptaddress,twocolumn,showkeys,preprintnumbers,floats,floatfix]{revtex4}

\usepackage{graphicx}
\usepackage{amsmath}

\begin{document}

\title{Observation of the single-electron regime in a highly tunable silicon quantum dot}

\author{W. H. Lim}
\email[These authors contributed equally to this work]{}
\affiliation{Centre of Excellence for Quantum Computer Technology, School of Electrical Engineering \& Telecommunications, The University of New South Wales, Sydney 2052, Australia}
\author{F. A. Zwanenburg}
\email[These authors contributed equally to this work]{}
\affiliation{Centre of Excellence for Quantum Computer Technology, School of Electrical Engineering \& Telecommunications, The University of New South Wales, Sydney 2052, Australia}
\email[These authors contributed equally to this work]{}
\author{H. Huebl}
\email[Present address: Walther-Meissner-Institut, Bayerische Akademie der Wissenschaften,
Walther-Meissner-Str. 8, 85748 Garching, Germany]{}
\affiliation{Centre of Excellence for Quantum Computer Technology, School of Electrical Engineering \& Telecommunications, The University of New South Wales, Sydney 2052, Australia}
\author{M. M\"{o}tt\"{o}nen}
\affiliation{Centre of Excellence for Quantum Computer Technology, School of Electrical Engineering \& Telecommunications, The University of New South Wales, Sydney 2052, Australia}
\affiliation{Department of Applied Physics/COMP and Low Temperature Laboratory, Helsinki University of Technology, P.O. Box 5100, FI-02015 TKK, Finland}
\author{K. W. Chan}
\affiliation{Centre of Excellence for Quantum Computer Technology, School of Electrical Engineering \& Telecommunications, The University of New South Wales, Sydney 2052, Australia}
\author{A. Morello}
\affiliation{Centre of Excellence for Quantum Computer Technology, School of Electrical Engineering \& Telecommunications, The University of New South Wales, Sydney 2052, Australia}
\author{A. S. Dzurak}
\affiliation{Centre of Excellence for Quantum Computer Technology, School of Electrical Engineering \& Telecommunications, The University of New South Wales, Sydney 2052, Australia}

\date{\today}

\begin{abstract}
We report on low-temperature electronic transport measurements of a silicon metal-oxide-semiconductor quantum dot, with independent gate control of electron densities in the leads and the quantum dot island. This architecture allows the dot energy levels to be probed without affecting the electron density in the leads, and vice versa. Appropriate gate biasing enables the dot occupancy to be reduced to the single-electron level, as evidenced by magnetospectroscopy measurements of the ground state of the first two charge transitions. Independent gate control of the electron reservoirs also enables discrimination between excited states of the dot and density of states modulations in the leads.

\end{abstract}

\pacs{71.55.-i, 73.20.-r, 76.30.-v, 84.40.Az, 85.40.Ry}

\keywords{quantum dot, silicon, single-electron, density of states modulations}

\maketitle

The ability to confine a single electron in a semiconductor quantum dot is an indispensable criterion for the implementation of quantum logic gates based on electrostatically confined electron spins, as proposed by Loss and DiVincenzo~\cite{Loss1998}. Most of the pioneering work on single-electron quantum dots has been done on devices based on GaAs/AlGaAs heterostructures~\cite{Elzerman2003,Elzerman2004}, in which the spin degree of freedom suffers from strong decoherence due to the presence of nuclear spins in the substrate. Therefore the group-IV semiconductors, which can be made essentially nuclear-spin free, constitute an appealing alternative for the fabrication of single-electron dots. The single-electron regime has already been achieved in nanowires of silicon~\cite{Zwanenburg2009}, carbon nanotubes~\cite{Herrero2004,Kuemmeth2008,Churchill2009,Steele2009}, and heterostructures of Si/SiGe~\cite{Simmons2007}.

Tunable silicon quantum dots~\cite{Angus2007,Liu2008,Lim2009} in metal-oxide-semiconductor (MOS) structures have been realized recently. However, it is difficult to obtain a single-electron dot in these devices because the upper gate simultaneously controls the occupancy of the dot and the electron density of the accumulation layer in the leads. Therefore, the leads are depleted and eventually turned off as the number of electrons in the dot is decreased, which inhibits transport experiments down to a single electron.

\begin{figure}[t]
\includegraphics[width=7.5cm]{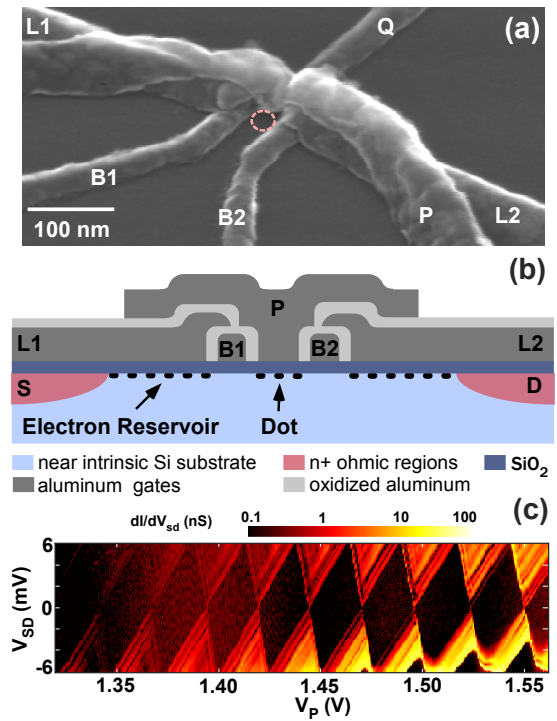}
\caption{(a) SEM image and (b) schematic cross section of a Si MOS quantum dot. The barrier gates, B1 and B2, control the tunnel coupling between the dot and the reservoirs. The lead gates, L1 and L2, control the electron density of the accumulation layer in the leads, while the plunger gate, P, independently controls the electron occupancy in the dot. The additional gate, Q, was not used in this experiment. (c) Stability diagram of the device in the many-electron regime for $V_{\textrm{L1}}$=$V_{\textrm{L2}}$=2.5~V, $V_{\textrm{B1}}$=0.568~V, and $V_{\textrm{B2}}$=0.492~V. By decreasing $V_{\textrm{P}}$, the tunnel barriers become more opaque.}
\label{fig1}
\end{figure}

In this letter we introduce a silicon quantum dot with independent control over the electron density in the island and the leads. This structure has overcome two key engineering challenges: (i) precision alignment between gate layers; and (ii) electrical isolation between gates to prevent leakage. By using a plunger gate, we can tune the energy levels in the dot. Additionally, the lead gates of the device enable us to vary the electron density in the reservoirs without affecting the dot, allowing for discrimination between excited states of the dot and density of states (DOS) modulations in the leads. With this multi-gate structure it is possible to reduce the occupancy of the quantum dot to a single electron, while still having a large electron density in the source$-$drain reservoirs to enable transport measurements. Magnetospectroscopy of the ground state of the first two charge transitions reveals that the first two electrons to occupy the dot are both spin-down.

The quantum dot structures investigated in this work were fabricated on a high-resistivity ($\rho >$ 10~k$\Omega$~cm at 300 K) silicon substrate. Standard Si microfabrication techniques were employed to create the thin gate-oxide, field-oxide and ohmic contacts. Electron-beam lithography (EBL), thermal evaporation and lift-off processes were used to pattern the aluminum gates above the SiO$_2$. The electrical isolation between gate levels was achieved by oxidizing the aluminum gates to create a 5-nm-thick aluminum oxide layer. Most of these fabrication steps have been described fully elsewhere~\cite{Angus2007} for previous Si MOS quantum dot structures. The devices discussed here also include a plunger gate, which is electrically isolated from the lower gates by aluminum oxide and patterned using EBL with an alignment accuracy of $\sim$20~nm between gate levels. Overall, we measured 12 devices; each with 5 gate electrodes at 4~K. Of the 60 gates tested, we found gate$-$substrate leakage in 10 gates, with 4 complete devices operating optimally. Subsequently, 2 quantum dot devices were cooled to 50~mK, both showing single-dot behavior.

Figures 1(a) and 1(b) show a scanning electron microscope (SEM) image and a schematic cross-section of a single-dot device, respectively. The lowest layer of gates are barrier gates (B1 and B2) with width 30~nm and separation 30~nm. The second layer of gates defines the source$-$drain leads (L1 and L2), which are patterned to overlap partially with the barrier gates, extending to the phosphorus in-diffused $n^+$ source and drain contacts. The plunger gate (P), which extends over the barrier gates, lead gates and the dot island, is $\sim$50~nm wide and $\sim$120~nm thick. The barrier gates define the dot spatially and control the tunnel coupling; the lead gates induce the electron accumulation layers that act as source$-$drain reservoirs; and the plunger gate controls the electron occupancy of the dot. The lithographic size of the quantum dot is estimated to be $\sim$30$\times$50~nm$^2$. This multi-gated structure provides excellent flexibility for tuning the barrier transparency and the quantized energy levels of the dot independently.

Electrical transport measurements were performed in a dilution refrigerator at a base temperature of $\sim$50 mK. An ac source$-$drain excitation voltage $V_{\textrm{sd}}$ typically below 100~$\mu$V at a modulation frequency of 13 Hz was applied to the device, in combination with a dc voltage $V_{\textrm{SD}}$. Standard lock-in techniques were used to monitor the differential conductance $dI/dV_{\textrm{sd}}$ while the source$-$drain dc current $I_{\textrm{SD}}$ was measured with a digital multimeter.

Figure 1(c) shows bias spectroscopy data with roughly 40 electrons on the dot. At high plunger gate voltage $V_{\textrm{P}}$, the device readily allows the creation of a single quantum dot, as evidenced by the stable and regular Coulomb diamonds. The charging energy $E_{\textrm{C}}=6$~meV in this regime, yielding a total capacitance of the quantum dot $C_{\Sigma}=e^2/E_{\textrm{C}}\approx30$~aF. From the period of Coulomb oscillations, we extract the plunger-gate-to-dot capacitance $C_{\textrm{P}}=e/\Delta V_{\textrm{P}}\approx6.5$~aF. The number of electrons can be reduced by decreasing $V_{\textrm{P}}$, but the tunnel barriers become less transparent. Due to the tunability of this device we are able to counteract the loss of barrier transparency by increasing $V_{\textrm{B1, B2}}$, which results in a measurable current through the dot down to its last electron.

\begin{figure}[t]
\includegraphics[width=8.5cm]{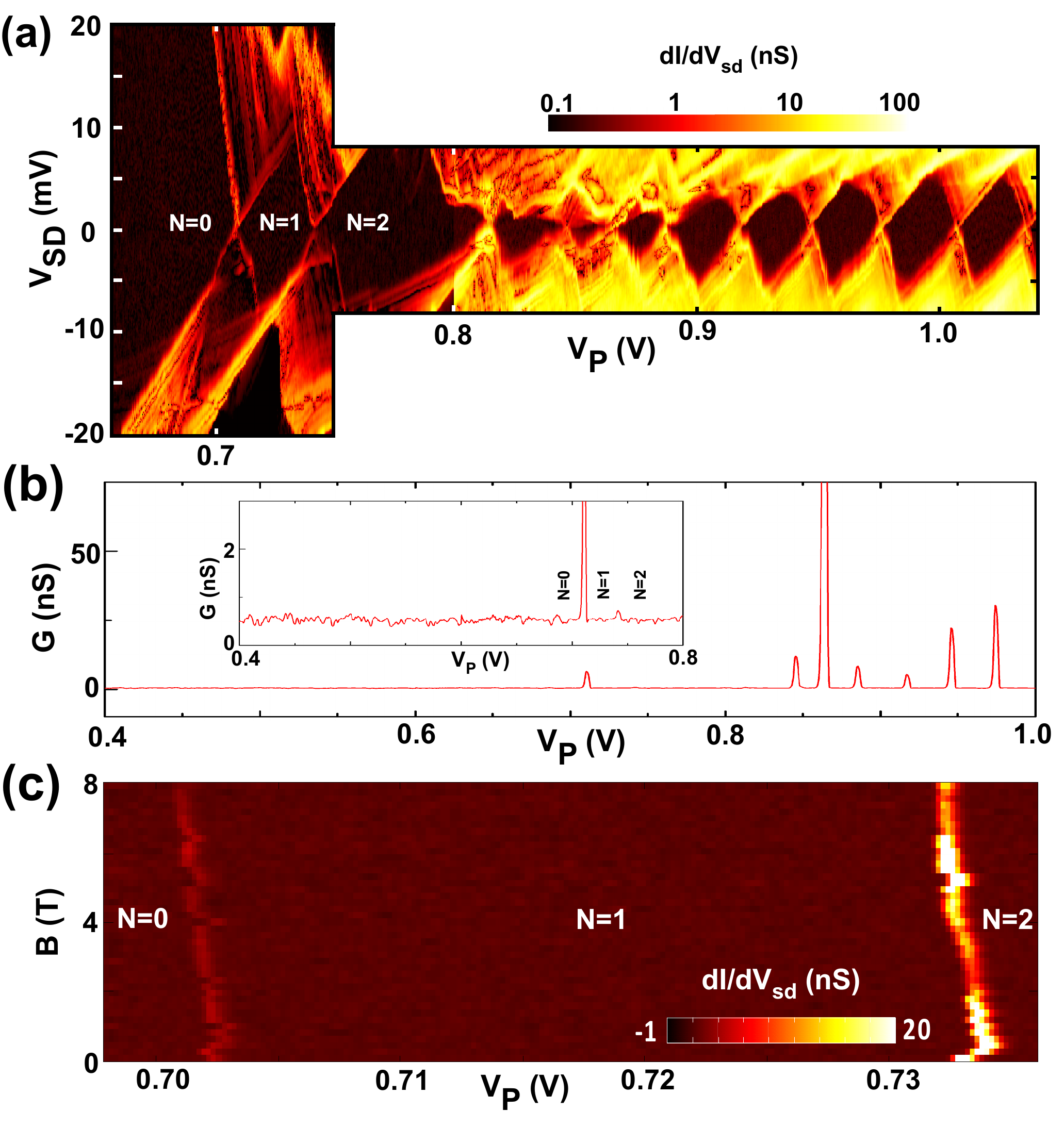}
\caption{(a) Stability diagram of the device in the few-electron regime for $V_{\textrm{L1}}$=$V_{\textrm{L2}}$=2.6~V, $V_{\textrm{B1}}$=0.485~V, and $V_{\textrm{B2}}$=0.660~V. The last diamond opens up completely, which is a strong indication that we observe the last electron tunneling out of the dot. Distortions from an unintentional dot are visible between $V_{\textrm{P}}$=800 and 900~mV. (b) The Coulomb peak heights of the last 10 electrons are typically 25~nS, and the conductance of the last peak is 12~nS. At lower $V_{\textrm{P}}$, there is no observable peak above the noise floor. (c) Magnetospectroscopy of the ground states of the first two electrons for $V_{\textrm{L1}}$=$V_{\textrm{L2}}$=2.45~V, $V_{\textrm{B1}}$=0.520~V, and $V_{\textrm{B2}}$=0.680~V. Evolution of the 0$-$1 and the 1$-$2 Coulomb peaks with magnetic field shows that two spin-down electrons enter the dot.}
\label{fig2}
\end{figure}

A stability diagram of the quantum dot in the few-electron regime is depicted in Fig.~2(a). After the last two charge transitions, the diamond edges open up entirely to a source$-$drain voltage $|V_{\textrm{SD}}|>20$~mV, a strong indication that the last electron has been depleted from the dot. We observe significant distortions (especially between $V_{\textrm{P}}$=0.8 and 0.9~V) owing to an unintentional parallel dot. This undesired effect has also masked the visibility of excited states at those charge transitions. In Figure 2(b), we plot the conductance as a function of plunger gate voltage at zero bias ($V_{\textrm{SD}}=0$). The Coulomb peak heights of the last 10 electrons are typically 25~nS [besides one exceptionally high peak ($\sim$350~nS) due to distortions from the accidental dot at $V_{\textrm{P}}$=0.86~V], and the conductance of the last peak is 12~nS. At lower $V_{\textrm{P}}$, there is no observable peak above the noise floor of 0.5~nS down to $V_{\textrm{P}}$=400~mV, nor do we observe any conductance above the noise floor at $V_{\textrm{SD}}$=$\pm$20~mV, providing strong evidence that the last electron has been depleted from the dot. In a further extension of this architecture, we plan to integrate a charge sensor to provide an additional monitor of the electron occupancy, as demonstrated for other quantum dot systems~\cite{Chan2006,Hu2007,Simmons2007}.

Magnetospectroscopy of the first two charge transitions has been performed by sweeping the plunger gate voltage $V_{\textrm{P}}$ while stepping the in-plane magnetic field, $B_\parallel$ [Figure 2(c)]. The Coulomb peak positions of the 0$-$1 and 1$-$2 transition both move towards less positive $V_{\textrm{P}}$, indicating two spin-down electrons entering the dot. In general, for a quantum dot having a non-degenerate orbital ground state, the lowest-energy two-electron state is a spin singlet. However, because of the valley degeneracy of silicon, one can have orthogonal states that are very close in energy. A small magnetic field can lower the energy of the spin triplet - where the two electrons are in different valley states - with respect to the singlet~\cite{Hada2003}. Therefore, we may interpret the results in Fig. 2 as a signature of small valley-orbit splitting. A linear fit through the positions yields $\Delta V_{\textrm{P}}=(703-0.196\times B_\parallel$)~mV for the 0$-$1 transition; and $\Delta V_{\textrm{P}}=(734-0.217\times B_\parallel$)~mV for the 1$-$2 transition. Each Coulomb peak moves by half the Zeeman energy $E_{\textrm{Z}}$=$|g|\mu_BB_\parallel$, where $\mu_B=58~\mu$eV/T is the Bohr magneton. Conversion of the slopes to energies using $\alpha$ (here, $\alpha \sim$0.29)  yields $g$-factors of 1.96$\pm$0.20 and 2.17$\pm$0.18 for the 0$-$1 and the 1$-$2 transition, respectively, consistent with electrons in bulk silicon.

\begin{figure}[t]
\includegraphics[width=9.0cm]{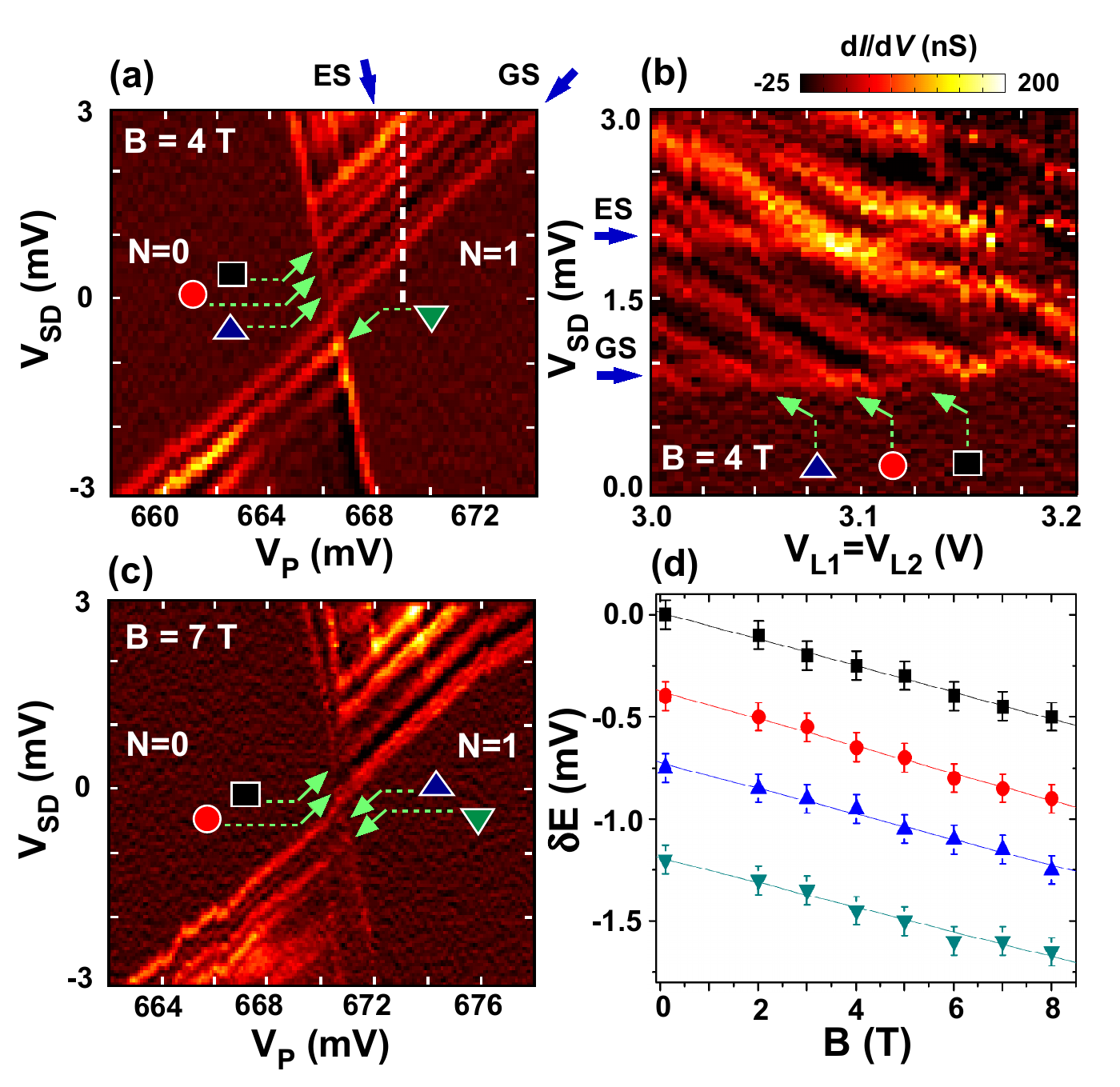}
\caption{Bias spectroscopy of the 0$-$1 transition of the dot at (a) B=4~T and (c) 7~T. (b) Plot of conductance as a function of $V_{\textrm{SD}}$ and $V_{\textrm{L1, L2}}$, corresponding to the dashed line in (a). In order to maintain the position of the charge transition while increasing $V_{\textrm{L1, L2}}$, we compensate by reducing $V_{\textrm{P}}$. Conductance peaks move down in energy with increasing $V_{\textrm{L1, L2}}$, which controls the Fermi level in the leads, whereas the orbital excited state and ground state peaks (blue arrows) stay constant. (d) Energy shift, $\delta E$, as a function of magnetic field is extracted from the conductance peaks (green arrows) shown in (a) and (c).}
\label{fig3}
\end{figure}

Lines of increased conductance in the stability diagrams can be attributed to e.g. orbital excited states in the dot or DOS modulations in the source$-$drain leads~\cite{Kouwenhoven1997}. Very recently, an efficient technique to probe and control the reservoir density of states has been developed~\cite{Mottonen2009}. Here, we use this method to pinpoint the origin of the resonant tunneling features in our device. Figures 3(a) and 3(c) show bias spectroscopy data for the 0$-$1 transition of the dot at $B_\parallel$=4~T and 7~T. The shift of the charge transitions in gate space compared to Fig.~2 is due to thermal cycling. Pronounced conductance peaks spaced by roughly 0.3~meV end on the Coulomb blockaded regions. In Figure 3(b), we show the conductance as a function of the source$-$drain voltage $V_{\textrm{SD}}$ and the lead voltage $V_{\textrm{L1, L2}}$. We compensate with the plunger gate voltage $V_{\textrm{P}}$ to remain at the same distance from the charge transition, corresponding to the dashed line trace in Fig.~3(a). The conductance peaks (green arrows) shift down in energy roughly linearly with increasing voltage on the lead gates, which controls the Fermi level of the accumulation layer in the leads. We note that the ground state (GS) and most likely an excited state (ES) of the dot (blue arrows) are independent of the lead voltage.

By following the conductance peaks (green arrows) in the stability diagrams systematically, we plot the energy shift $\delta E$ versus magnetic field in Fig.~3(d). The fitted line indicates a slope of $-\frac{1}{2}g\mu_B/e$, that is half the Zeeman energy, in agreement with the interpretation that the conductance peaks are due to DOS modulations in the leads. In a magnetic field, the orbital GS of the dot splits into spin-down and spin-up states by the Zeeman energy. This shifts the spin-down state down by $\frac{1}{2}g\mu_BB$, as observed in the magnetospectroscopy of the first charge transition in Fig.~2(c). In the stability diagrams of Figure~3(a) and 3(c), the DOS modulations remain at the same position for different magnetic fields. As a result the resonances corresponding to DOS modulations shift by $\frac{1}{2}g\mu_BB$ with reference to the edges of the Coulomb diamonds. A more detailed study of this quasi-one-dimensional DOS in narrow MOS structures can be found in~\cite{Mottonen2009}. The results in Figure~3 show that we can discriminate between excited states of the dot and the DOS in the leads.

In summary, we have fabricated MOS quantum dot devices with independent control of the electron densities in the leads and the quantum dot island. This design enabled us to observe the last electron leaving the quantum dot. The first two electrons entering the empty dot are both spin-down, most likely filling two nearly-degenerate valley states. Finally, we showed that most conductance peaks in the bias spectroscopy correspond to resonances with the density of states in the leads. This quantum dot design provides great promise for future experiments on spin and valley physics in silicon.

The authors thank D.~Barber and R. P.~Starrett for technical support in the National Magnet Laboratory at University of New South Wales, and M. A. Eriksson for helpful discussions. M. M. acknowledges Emil Aaltonen Foundation for financial support. This work was supported by the Australian Research Council, the Australian Government, and by the U. S. National Security Agency and U. S. Army Research Office (under Contract No. W911NF-08-1-0527).

\end{document}